\documentclass[pra,twocolumn,showpacs,superscriptaddress]{revtex4}

\usepackage{graphicx}
\usepackage{amsfonts}
\usepackage{amsmath}
\usepackage{amssymb}

\begin{document}

\title{Isotope effects in the harmonic response from hydrogenlike muonic atoms in strong laser fields}

\author{Atif Shahbaz}
\altaffiliation{Permanent address: Department of Physics, GC University, 54000 Lahore, Pakistan}
\affiliation{Max-Planck-Institut f\"{u}r Kernphysik, Saupfercheckweg 1, 69117 Heidelberg, Germany}
\author{Thomas J. B\"{u}rvenich}
\affiliation{Frankfurt Institute for Advanced Studies, Johann Wolfgang Goethe University, Ruth-Moufang-Str.~1, 60438 Frankfurt am Main, Germany}
\author{Carsten M\"{u}ller}
\email[Corresponding author: ]{c.mueller@mpi-k.de}
\affiliation{Max-Planck-Institut f\"{u}r Kernphysik, Saupfercheckweg 1, 69117 Heidelberg, Germany}

\date{\today}

\begin{abstract}
High-harmonic generation from hydrogenlike muonic atoms exposed to ultraintense high-frequency laser fields is calculated. Systems of low nuclear charge number $Z$ are considered where a nonrelativistic description applies. By comparing the radiative response for different isotopes we demonstrate characteristic signatures of the finite nuclear mass and size in the harmonic spectra. In particular, for $Z>1$, an effective muon charge appears in the Schr\"odinger equation for the relative particle motion, which influences the position of the harmonic cutoff. Cutoff energies in the MeV domain can be achieved, offering prospects for the generation of ultrashort coherent $\gamma$-ray pulses.
\end{abstract}
\maketitle

%
\section{Introduction}

One of the most successful and accurate methods to probe nuclear properties employs muonic atoms \cite{EiGr1970,MuRev}. Due to the small Bohr radius of these exotic atoms, the muonic wave function has a large overlap with the binding nucleus. Precision measurements of muonic transitions to deeply bound states can therefore reveal nuclear structure information such as finite size, deformation, surface thickness, and polarization. The first x-ray spectroscopy of muonic atoms was performed in 1953 using a 4-meter cyclotron \cite{MuFirst}. Today, large-scale facilities like TRIUMF (Vancouver, Canada) or PSI (Villingen, Switzerland) exist which are specialized in the efficient generation of muons and muonic atoms \cite{TRIUMF}. New developments aim at the production of radioactive muonic isotopes for conducting spectroscopic studies on unstable nuclear species \cite{MuRadioactive}. Muons bound in atoms are also able to catalyze nuclear fission \cite{Ga1980} and fusion \cite{BrKaCoLe1989} reactions.

On a different front, the field of laser-nuclear physics is emerging \cite{ScMaBe2006}. While lasers have always represented important tools for nuclear spectroscopy \cite{Ne2002}, in recent years their role is changing qualitatively and growing because of the tremendous progress in high-power laser technology. The interaction of intense short laser pulses ($I\sim 10^{18}-10^{20}$\,W/cm$^{2}$) with matter can produce highly energetic electrons, protons and photons (e.g., via bremsstrahlung). In pioneering experiments, this has led to the observation of laser-induced nuclear fission \cite{LeCo2000}, nuclear fusion \cite{Di1997Di1999}, and neutron production in nuclear reactions \cite{Um2003LeMcSi2003}. Advanced laser sources might also pave the way to nuclear quantum optics \cite{1BuEvKe2006} and coherent $\gamma$-spectroscopy using ultrashort pulses \cite{GRASER,Klaiber,yocto}.

In light of this, the combination of muonic atoms with intense laser fields opens promising perspectives. Contrary to the traditional spectroscopy of muon transitions between stationary bound states, the exposure of a muonic atom to a strong laser field renders the problem explicitly time-dependent and the muon, thus, a \textit{dynamic} nuclear probe. In this setup, the muon is coherently driven across the nucleus which, for example, gives rise to the emission of radiation and, in general, allows for time-resolved studies on a femtosecond scale. The information on the nucleus gained by laser-assistance can in principle complement the knowledge obtained from the usual field-free spectroscopy of muonic atoms. 

Against this background, we have recently considered the process of high-harmonic generation (HHG) from strongly laser-driven muonic hydrogen and deuterium atoms \cite{ShMuStBuKe2007}. The process of HHG represents a frequency up-conversion of the applied laser frequency due to a nonlinear coupling of the atom with the driving external field (see \cite{BeMi2002,Salamin,Bauer,Scrinzi2006,moreHHG} for recent reviews). It can be understood within a three-step model, where the bound lepton is liberated from the atom by tunneling ionization, propagates in the laser field, and finally recombines with the core, returning its kinetic energy upon photoemission. By way of a comparative study it was demonstrated that the harmonic response from muonic hydrogen isotopes is sensitive to the nuclear mass and size \cite{ShMuStBuKe2007}. This shows that muonic atoms subject to strong laser fields can reveal information on nuclear degrees of freedom. Muonic deuterium molecules in superintense laser fields represent another interesting example towards this combined effort, where field-induced modifications of muon-catalyzed fusion have been investigated \cite{ChBaCo2004}. Muonic hydrogen atoms have moreover been studied as systems which could allow for observation of the Unruh effect \cite{KaSc2005}.

In this paper we extend our previous study on HHG \cite{ShMuStBuKe2007} to hydrogenlike muonic atoms (ions) with nuclear charge number $Z\ge 1$. To this end, the time-dependent Schr\"odinger equation in the presence of the binding nucleus and a few-cycle laser pulse is solved on a numerical grid in one spatial dimension. From the resulting dipole acceleration the HHG spectrum is extracted and compared for different isotopes. Characteristic effects arising from the finite nuclear mass and size are revealed. Moreover, in the case $Z>1$ the laser-driven particle dynamics becomes more complex because the center-of-mass of the atomic constituents does not stay at rest any longer. As a result, an effective muon charge appears in the Schr\"odinger equation for the relative motion, which affects the harmonic cutoff position. The cutoff energies achievable with muonic atoms in the nonrelativistic domain of interaction are very large, reaching several MeV. This holds in principle prospects for the production of coherent $\gamma$-ray pulses of ultrashort duration.

Since muonic atoms are tightly bound systems, laser fields of extraordinary field strength and photon energy are required to influence the muon motion. In the ground state of muonic hydrogen, for example, the muon is bound by 2.5\,keV and experiences a binding Coulomb field strength of $1.8\times10^{14}$ V/cm corresponding to the field intensity $4.2\times10^{25}$ W/cm$^{2}$. A comparison of these numbers with the parameters of the most advanced present-day and near-future laser sources is useful. In the range of optical and near-infrared frequencies ($\hslash \omega \sim 1$\,eV), the highest intensity presently attainable is $\sim10^{22}$\,W/cm$^{2}$ \cite{Yanovsky} and the next generation of high-power lasers aims at intensities of $10^{23}$\,W/cm$^{2}$ and beyond \cite{MoTaBu2006}. In the VUV frequency domain ($\hslash \omega \sim 10-100$\,eV) a maximum intensity of $\sim 10^{17}$\,W/cm$^{2}$ has been attained with a free-electron laser at the FLASH facility (DESY, Germany) \cite{web2006}. The Linac Coherent Light Source (SLAC, Stanford) has recently entered the frequency domain $\hslash \omega \sim 1$\,keV \cite{LCLS}. Near-future upgrades of such machines are planned to produce brillant x-ray beams ($\hslash \omega \sim 10$\,keV) with peak intensities close to $10^{20}$\,W/cm$^{2}$. There are also efforts to generate ultrashort, high-frequency radiation ($\hslash \omega \sim 10-1000$\,eV) from plasma surface harmonics where considerably higher intensities might be reachable due to a high conversion efficiency \cite{Ts2006}. With these novel sources of intense coherent radiation it will become possible to influence the quantum dynamics of light muonic atoms with nuclear charge numbers $Z\lesssim 10$. Note that in principle the laser field strengths and frequencies experienced by the atoms can be enhanced further when a relativistic atomic beam is employed instead of a fixed target \cite{Ciprian,Ug2006}.

As to their lifetime, we point out that light muonic atoms and molecules may be regarded as quasistable systems on the ultrashort time-scales of strong laser pulses ($\tau\sim {\rm fs}$--ps), since their life time is determined by the free muon life time of 2.2\,$\mu$s. For the field parameters assumed in this paper, the influence of the external laser field on the muon decay is immaterial as well \cite{Dicus}. In deeply bound states of heavy atoms, the muon life time can be reduced due to absorption by the nucleus to $\sim 10^{-8}$\,s which still exceeds typical laser pulse durations by orders of magnitude.

We organize the paper as follows: Section II deals with the theoretical framework in which the separation into center-of-mass and relative coordinates of the two-body Schr\"{o}dinger equation for a hydrogenlike muonic atom in a laser field is performed. We also give here a scaling transformation between ordinary and muonic atoms, as well as suitable model potentials which allow us to incorporate the nuclear mass and size. Section III has been reserved for the presentation of our numerical results and their discussion. While Sec.\,III.A compiles the harmonic cutoff energies available from different low-$Z$ muonic atoms, Sects.\,III.B and III.C show a series of calculations devoted to the impact of the nuclear mass and size on the HHG spectra. A comparison of the nuclear signatures predicted for muonic atoms with those to be expected in highly-charged electronic ions is undertaken in Sec.\,III.D. The conclusion of the paper is given in Sec.\,IV.

%
\section{Theoretical framework}
%
\subsection{Separation of relative and center-of-mass motion}
We consider the nonrelativistic quantum dynamics of an initially bound muon in a few-cycle laser pulse described by the time-dependent Schr\"{o}dinger equation (TDSE). For our laser parameters of interest, we may ignore the space dependence of the laser field (dipole approximation), treating it as a purely time-dependent electric field. Due to the large muon mass, the atomic nucleus cannot be considered as infinitely heavy. We therefore start from the two-particle TDSE written in the length gauge as:

\begin{align}
\label{TDSE2}
i\hslash\dfrac{\partial}{\partial t}\psi \left(\textbf{x}_{\mu},\textbf{x}_{n};t\right)= & \bigg[\dfrac{\textbf{p}_{\mu}^{2}}{2m_{\mu}}+\dfrac{\textbf{p}_{n}^{2}}{2m_{n}}+e\textbf{x}_{\mu}\cdot\textbf{E}(t)\nonumber\\
&\!\!\!\!\!\!\!\!\!\!\!\!\!\!\!\!\! -Ze\textbf{x}_{n}\cdot\textbf{E}(t)
  +V\left(\vert\textbf{x}_{\mu}-\textbf{x}_{n}\vert\right)\bigg]\psi \left(\textbf{x}_{\mu},\textbf{x}_{n};t\right)
\end{align}
where $m_{\mu}$ and $m_{n}$ are the muonic and nuclear masses, $\textbf{x}_{\mu}$ and $\textbf{x}_{n}$ are the coordinate vectors for the muon and the nucleus, and $\textbf{p}_{\mu}=-i\hslash\partial/\partial \textbf{x}_{\mu}$ and $\textbf{p}_{n}=-i\hslash\partial/\partial \textbf{x}_{n}$ the corresponding momentum operators, respectively. Besides, the nuclear charge number is $Z$, the elementary charge unit $e$, the binding potential $V\left(\vert\textbf{x}_{\mu}-\textbf{x}_{n}\vert\right)$, and the laser electric field $\textbf{E}(t)$ which oscillates with angular frequency $\omega$. 

The application of the dipole approximation in Eq.\,(\ref{TDSE2}) considerably simplifies the problem as it renders the  Schr\"odinger equation (\ref{TDSE2}) for the muon-nucleus two-body system separable into relative and center-of-mass motion. By introducing relative and center-of-mass coordinates $\textbf{x}=\textbf{x}_\mu-\textbf{x}_n$ and $\textbf{X}=(m_\mu\textbf{x}_\mu + m_n\textbf{x}_n)/M$, respectively, with the total mass $M=m_{\mu}+ m_{n}$, one finds that the evolution of the center-of-mass is described by
\begin{equation}
\label{CMSE}
i\hslash\dfrac{\partial}{\partial t}\Psi\left( \textbf{X},t\right)=\left[\dfrac{\textbf{P}^{2}}{2M}-(Z-1)e\textbf{X}\cdot\textbf{E}\left(t\right)\right]\Psi\left( \textbf{X},t\right)
\end{equation}
with $\textbf{P}=-i\hslash\partial/\partial \textbf{X}$. Equation\,(\ref{CMSE}) is the non-relativistic Volkov equation for a particle of charge $(Z-1)e$ and mass $M$ in the presence of a laser field. As a consequence, the center-of-mass motion does not emit higher harmonic frequencies and may therefore be ignored in the following. Note that in the special case $Z=1$ (i.e., hydrogen isotopes) the center-of-mass moves freely, while the laser field only couples to the relative coordinate \cite{RuPlRoBe2006}.

The relative motion is governed by (see also \cite{ChBaCo2004,MaLa1999})
\begin{equation}
\label{RSE}
i\hslash\dfrac{\partial}{\partial t}\psi \left(\textbf{x},t\right)=\left[\dfrac{\textbf{p}^{2}}{2m_{r}}+q_{e}\textbf{x}\cdot\textbf{E}(t)+V\left(\textbf{x}\right)\right]\psi \left(\textbf{x},t\right)
\end{equation}
with the reduced mass $m_{r}=m_{\mu} m_{n}/M$, the relative momentum $\textbf{p}=-i\hslash\partial/\partial \textbf{x}$, and the effective charge
\begin{equation}
\label{qeff}
q_{e} = m_{r}\left(\dfrac{Z}{m_{n}}+\dfrac{1}{m_{\mu}}\right)e.
\end{equation}
In the special case $Z=1$, the effective charge reduces to $q_e=e$, whereas $q_e>e$ holds for atomic numbers $Z>1$.

Formally, Eq.\,(\ref{RSE}) is the Schr\"{o}dinger equation for a single particle of charge $-q_e$ and mass $m_{r}$ in the presence of a nucleus and a laser field. The accelerated motion of the relative coordinate in the combined external fields therefore gives rise to the emission of higher harmonics. 
We note that, in physical terms, the relative coordinate accounts for the fact that both the nucleus and the muon oscillate in the laser field with different amplitudes in opposite directions (see also Fig.\,2 in \cite{ShMuStBuKe2007}).

We point out that the effective charge in Eq.\,(\ref{qeff}) has already been derived in Ref.\,\cite{Reiss}. It was also shown there that the two-body TDSE in a laser field separates straightforwardly in the velocity gauge. It is interesting to note that, in contrast, the separability in the length gauge is a more subtle issue because the operations of performing gauge transformations and dipole approximations are not commutative  \cite{Reiss,Bandrauk}. As a consequence, cross terms appear in the exact version of the two-body TDSE in the length gauge which, strictly speaking, prevent the equation from being separable. The cross terms typically become important at the borderline to the relativistic regime when the value of the relativistic field parameter [see Eq.\,(\ref{RP})] approaches unity \cite{Reiss}. For the nonrelativistic laser parameters applied in the present study, however, these terms are very small and have therefore been neglected in Eq.\,(\ref{TDSE2}).
%
%
\subsection{Scaling considerations}

The form of Eq.\,(\ref{RSE}) is equivalent to the Schr\"odinger equation for an ordinary hydrogen atom (i.e., an electron bound to an infinitely heavy proton) in a laser field. 
The relation can be made explicit by virtue of a general method known as scaling transformation. For the special case of a laser-driven atom, the scaling procedure is nicely explained in \cite{MaLa1999}. Suppose, that we have a hydrogenlike muonic system of nuclear charge $Z$ on the one side and an ordinary hydrogen atom on the other side. 
We introduce an electronic coordinate vector $\textbf{x}_e$ and time $t_e$ and relate them to the muonic coordinate $\textbf{x}$ and time $t$ of Eq.\,(\ref{RSE}) according to 
\begin{eqnarray}
\label{xete}
\textbf{x}_{e}=\dfrac{Z}{\rho} \textbf{x}_{\mu}\ ;\hspace{0.5cm}t_{e}=\dfrac{Z^{2}}{\rho}t_{\mu}
\end{eqnarray}
with the mass ratio $\rho \equiv m_{e}/m_{r}$ and the electron mass $m_{e}$. When rewritten in the scaled space and time, Eq.\,(\ref{RSE}) becomes
\begin{equation}
\label{hydrogenSE}
i\hslash \dfrac{\partial}{\partial t_{e}}\psi \left(\textbf{x}_{e},t_{e}\right)=\left[\dfrac{\textbf{p}_e^2}{2m_{e}} + e \textbf{x}_{e}\cdot\textbf{E}_{e}(t_e) + V(\textbf{x}_{e})\right]\psi \left(\textbf{x}_{e},t_{e}\right)
\end{equation}
with $\textbf{p}_e=-i\hslash\partial/\partial \textbf{x}_e$ and the scaled laser frequency and field strength
\begin{eqnarray}
\label{scaling}
\omega_e=\frac{\rho}{Z^2}\omega\ ; \hspace{0.5cm}\textbf{E}_e=\frac{q_e}{e}\frac{\rho^2}{Z^3}\textbf{E}\,.
\end{eqnarray}
This means that a muonic hydrogenlike atom in a laser field with parameters $E$ and $\omega$ behaves like an ordinary hydrogen atom in a field with $E_e$ and $\omega_e$ given by Eq.\,(\ref{scaling}), provided that the binding potential $V(x)$ arises from a pointlike nucleus. To give an example, the typical parameters of an intense Ti:Sapphire laser $\hslash\omega_e=1.5$\,eV, $E_e=2.7\times 10^8\,$V/cm ($10^{14}\,$W/cm$^2$) translate to a muonic helium atom as $\hslash\omega=1.2$\,keV, $E=9.1\times 10^{13}\,$V/cm ($1.1\times 10^{25}\,$W/cm$^2$). This comparison demonstrates that, despite the huge laser intensities applied in our computations, the laser-driven muon dynamics remains nonrelativistic due to the large muon mass.
Moreover, since the laser intensities required for HHG from muonic atoms are very large already for hydrogen isotopes and steeply increase with the nuclear charge, we restrict our consideration to muonic atoms with $Z\lesssim 10$. An advantage of these systems as compared to heavier ones is that the relative differences in mass and size among isotopes are larger for low-$Z$ atoms in the nuclear chart.

We emphasize that the scaling procedure does not account for nuclear properties like the finite nuclear size or the nuclear shape. Evidently, when the transition from, e.g., a muonic hydrogen atom to an ordinary hydrogen atom is performed, the proton radius is not to be length-scaled in accordance with Eq.\,(\ref{xete}) but remains fixed. As a consequence, for atomic systems where nuclear properties play a role, not all physical information can be obtained from the knowledge of the ordinary-atom case via scaling. In Sec.\,III.C below we show results which display the influence of the nuclear size on the process of HHG.
%
%
%
\subsection{1D approximation and model potentials}

Since we restrict ourselves to the consideration of the interaction regime where the dipole approximation applies, we can further simplify the problem by reducing the dimensionality. We only treat the muon motion in one dimension (1D) along the laser polarization axis. This is indicated, as in our case the 1D numerics still is a non-trivial task because of the fine grid spacing required to resolve the nuclear extension and the non-standard laser parameters employed. As regards ordinary atoms, the latter would correspond to intense fields in the mid-infrared, giving rise to large muon momenta and ponderomotive energies.
1D models are widely used in strong-field physics \cite{FaKoBeRo2002} as they retain the essential physical features of nonrelativistic laser-atom interaction for linearly polarized fields \cite{EbBook1992}. 

The main shortcoming of the 1D approach is the neglect of the muon's wave packet spreading in transversal direction. During one laser period $T=2\pi/\omega$, the spreading can be estimated as $\Delta r \sim \Delta v T$, where $\Delta v \approx (\hbar eE/\sqrt{2m_{r}^{3}I_{p}})^{1/2}$ is the velocity width at the exit of the potential barrier which the muon tunnels through \cite{Scrinzi2006} and $I_p$ denotes the atomic ionization potential. In an ultrastrong VUV field ($I=10^{23}$ W/cm$^{2}$, $\hbar \omega =60$ eV) we obtain $\Delta r \sim 10$ pm for muonic hydrogen, which is by a factor $\rho$ smaller than the spreading of an electron wave packet in an appropriately scaled infrared laser field [cf. Eq.\,(\ref{scaling})]. The wave packet spreading can substantially reduce the total  harmonic yield. We stress, however, that the goal of the present study is to reveal \textit{relative} differences between physical observables which typically are less sensitive to model assumptions than absolute numbers. 

In order to unveil the effects of the nuclear mass and size in the HHG spectra of muonic atoms we employ the following potentials in the 1D version of Eq.\,(\ref{RSE}):

\subsubsection{Soft-core potential}
Numerical calculations in reduced dimensionality usually employ a soft-core potential to describe the muon-nucleus interaction \cite{EbBook1992}. In this manner, the Coulomb singularity of a pointlike nucleus at the origin is avoided. We also apply a standard soft-core potential in our computations, which after appropriate scaling reads 
\begin{equation}
\label{SC}
V_{s}(x)=-\dfrac{Ze^{2}}{\sqrt{x^{2}+\left(\dfrac{\rho}{Z}\right)^{2}}}\,.
\end{equation}
It enables us to study the influence of the nuclear mass, which enters Eq.\,(\ref{SC}) via the reduced mass contained in the parameter $\rho$. The corresponding results are shown in Sec.\,III.B.

\subsubsection{Hard-core potential}
In order to describe the effect of the finite nuclear extension, the softcore potential (\ref{SC}) is not suitable. Instead, we apply for this purpose the nuclear drop model and consider the nucleus as a sphere of uniform charge density within the nuclear radius $R$. The corresponding potential is
\begin{equation}
\label{HC}
V_{h}(x)=\begin{cases} -\dfrac{Ze^{2}}{R}\left( \dfrac{3}{2}-\dfrac{x^{2}}{2R^{2}}\right)  &  \text{if $|x|\leq R$,}
\\
-\dfrac{Ze^{2}}{|x|} &\text{if $|x|>R$}
\end{cases}
\end{equation}
which explicitly takes the nuclear radius into account. We point out that in the limit $R\rightarrow0$, the binding energy of the lowest lying state of this potential becomes infinite and, thus, unphysical \cite{ScFa1994}. Following a well-established procedure \cite{ScFa1994,Ch1996,Gordon}, we therefore start our calculation from the first excited state which has the correct binding energy. By monitoring the projection onto the unphysical state during the time evolution, we take care that the occupation of this state always stays negligibly small.

%
\section{Results}
The TDSE (\ref{RSE}) for the relative motion has been solved numerically in one spatial dimension via the Crank-Nicolson time-propagation scheme. The laser field is always chosen as a 5-cycle pulse of trapezoidal envelope having one cycle for linear turn-on and one for turn-off. The HHG spectrum is obtained from a Fourier transformation of the dipole acceleration. 

\subsection{Maximum cutoff energies}

Since the conversion efficiency into high harmonics is rather low ($\sim 10^{-6}$), it is generally desirable to maximize the radiative signal strength. In our situation, the optimization is of particular importance as the target density of muonic atoms is low. A sizeable HHG signal requires efficient ionization on the one hand, as well as efficient recombination on the other hand. The former is guaranteed if the laser peak field strength lies just below the border of over-barrier ionization (OBI) where the Coulomb barrier is suppressed all the way to the bound energy level by the laser field \cite{Salamin}. From Eq.\,(\ref{RSE}) we obtain
\begin{equation}
\label{OBI}
E\lesssim E^{\rm OBI} = \dfrac{m_{r}^{2}c^{3}}{q_{e}\hslash}\,\dfrac{(\alpha Z)^{3}}{16}\,,
\end{equation}
with the fine-structure constant $\alpha \approx 1/137$. Efficient recollision is guaranteed if the magnetic drift along the laser propagation direction can be ignored, which limits the relativistic parameter to \cite{WaKeScBr2000,JOSA}
\begin{equation}
\label{RP}
\xi\equiv\dfrac{q_{e}E}{m_{r}c\omega}<\left(\dfrac{16\hslash \omega}{\sqrt{2m_{r}c^{2}I_{p}}} \right)^{1/3}.
\end{equation}
The condition (\ref{RP}) also confirms the applicability of the dipole approximation in Eq.\,(\ref{RSE}). 

The Eqs.\,(\ref{OBI}) and (\ref{RP}) above define a maximum laser intensity $I_{\rm max}\approx I^{\rm OBI}$ and a minimum laser frequency 
\begin{equation}
\label{wmin}
\omega_{\rm min} = \frac{m_rc^2}{16\hslash}(\alpha Z)^{5/2}
\end{equation}
which are still in accordance with the conditions imposed. At these laser parameters, the maximum harmonic cutoff energies are attained, while Eqs.\,(\ref{OBI}) and (\ref{RP}) guarantee an efficient ionization-recollision process. Note, however, that smaller driving frequencies generally lead to reduced harmonic signal strengths because of the more pronounced and unavoidable quantum wave-packet spreading \cite{MaFr2007,Sc2007,Burgdorfer} (for an exeption to this rule, see \cite{Frolov}). For muonic hydrogen the lowest frequency according to Eq.\,(\ref{wmin}) lies in the VUV range, $\hslash\omega_{\rm min}\approx27$\,eV, while the maximum field intensity is $I^{\rm OBI}\approx1.6\times10^{23}$ W/cm$^{2}$. At these values, the harmonic spectrum extends to a maximum energy of $\epsilon_{\rm max}\approx0.55$ MeV. For light muonic atoms with nuclear charge number $Z>1$, the achievable cutoff frequencies are even higher, reaching several MeVs. A summary is given in Table\,\ref{tableCutoff}. 

For comparison we note that the highest harmonic cutoff energy which has been attained experimentally with ordinary (helium) atoms, amounts to $\approx 1$\,keV \cite{Seres2005}; the corresponding harmonic order at the cutoff was $\epsilon_{\rm max}/\omega\approx 800$. Higher cutoff energies are difficult to achieve due to the detrimental effects of dephasing \cite{Justin} and electron drift motion; various schemes have been proposed to overcome this obstacle (see \cite{Salamin,JOSA,beyondDA} and references therein). Muonic atoms are advantageous in this respect since the large muon mass in principle allows for the generation of MeV harmonics in the dipole regime of interaction.

As a result, muonic atoms are promising candidates for the generation of hard x-rays or even $\gamma$-rays which might be employed to trigger photo-nuclear reactions.

\begin{table}[ht]
\centering
\begin{tabular}{c|c|c|c|c}
\hline
\ \ \ $Z$\ \ \ &\ \ \ \ \ $\hslash\omega_{\rm min}$\ \ \ \ \ &\ \ \ \ $\xi_{\rm min}$\ \ \ \ &\ \ \ \ $\xi_{\rm max}$\ \ \ \ &\ \ \ \ \ \ $\epsilon_{\rm max}$\ \ \ \ \ \ \\
\hline
 1 &  27\,eV  & 0.007 & 0.085 & 0.55 MeV\\
 2 & 170\,eV  & 0.015 & 0.12  & 1.1\,MeV\\
 4 & 960\,eV  & 0.03  & 0.17  & 2.2\,MeV\\
 10 & 9.5\,keV & 0.07 & 0.27  & 5.7\,MeV\\
\hline
\end{tabular}
\caption{Maximum HHG cutoff energies $\epsilon_{\rm max}$ achievable with hydrogenlike muonic atoms of nuclear charge number $Z$. The applied laser frequency $\omega_{\rm min}$ and intensity parameter $\xi_{\rm max}$ are chosen in accordance with Eqs.\,(\ref{OBI})-(\ref{wmin}) to allow for an efficient ionization-recollision process. $\xi_{\rm min}$ denotes the minimum intensity parameter leading to tunneling ionization \cite{Salamin}.}
\label{tableCutoff}
\end{table}

\subsection{Nuclear mass effects}

In this section we consider the effect on the HHG process stemming from a variation of the nuclear mass, assuming the nucleus as being point-like. We have solved the Schr\"odinger equation (\ref{RSE}) using the softcore potential (\ref{SC}). The latter depends on the nuclear mass via the reduced mass entering the parameter $\rho$. It is meaningful to consider light isotopes where the reduced mass $m_r$ is significantly different from the bare muon mass $m_\mu$. The nuclei chosen for the calculations in this and the following section are given in Table\,\ref{table}. In order to demonstrate the nuclear mass effect on the HHG process, isoptopes of large relative mass difference have been selected.

\begin{table}[ht]
\centering
\begin{tabular}{c|cc|cc}
\hline
Isotope\ &\ Mass $m_n$ (GeV$/c^2$) &\ Ref.\ \ \ &\ Size $R$ (fm) &\ Ref.\\
\hline
H        & 0.9383 & \cite{Ro1994} & 0.875 & \cite{MoTa2005}\\
D        & 1.8756 & \cite{Ro1994} & 2.139 & \cite{MoTa2005}\\
$^{3}$He & 2.8084 & \cite{Ro1994} & 1.9448 & \cite{An2004}\\
$^{4}$He & 3.7274 & \cite{Ro1994} & 1.6757 & \cite{An2004}\\
$^{6}$Li & 5.6016 & \cite{Ro1994} & 2.517 & \cite{Sa2006}\\
$^{9}$Li & 8.4069 & \cite{Ro1994} & 2.217 & \cite{Sa2006}\\
$^{20}$Ne & 18.493 & \cite{NaSa1999} & 3.0053 & \cite{An2004}\\
$^{23}$Ne & 21.277 & \cite{NaSa1999} & 2.9126 & \cite{An2004}\\
\hline
\end{tabular}
\caption{Nuclear masses and rms charge radii of various isotopes which have been used in the calculations. The half lives of $^9$Li and $^{23}$Ne amount to 178\,ms and 37\,s, respectively; the other nuclei are stable.}
\label{table}
\end{table}

\begin{figure}[b]
(a)\includegraphics*[width=0.62\linewidth,angle=270]{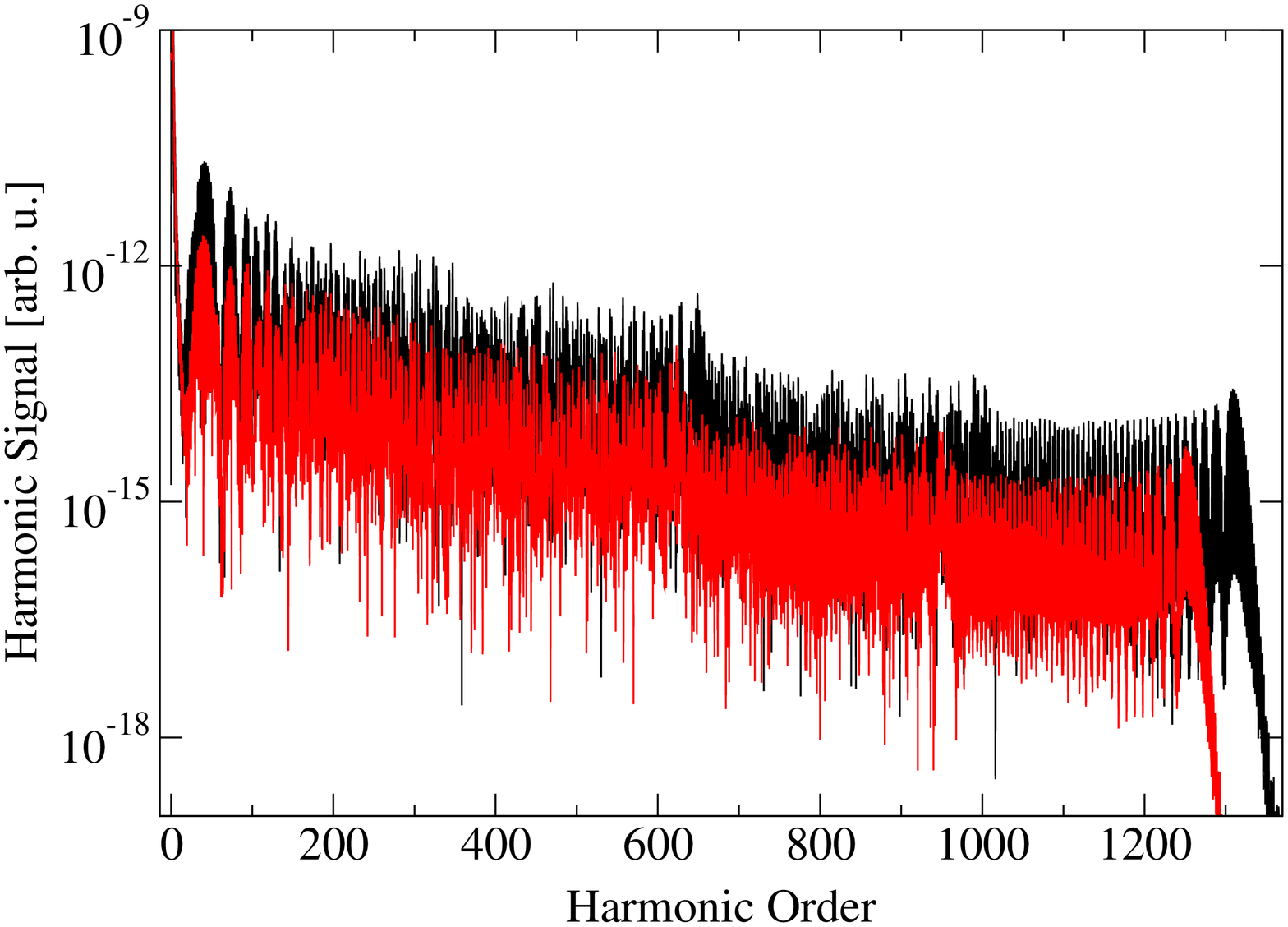}
(b)\includegraphics*[width=0.6\linewidth,angle=270]{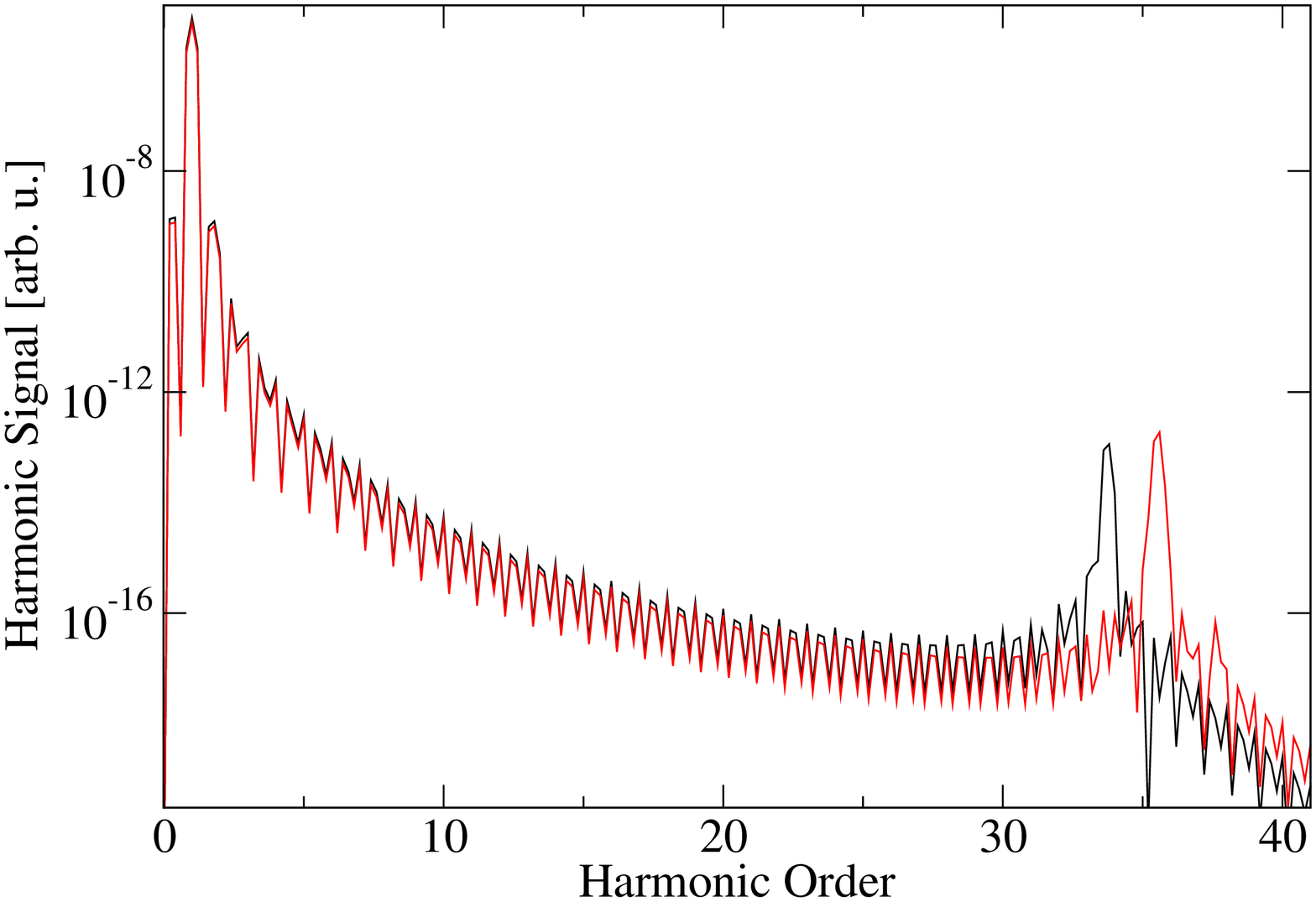}
(c)\includegraphics*[width=0.6\linewidth,angle=270]{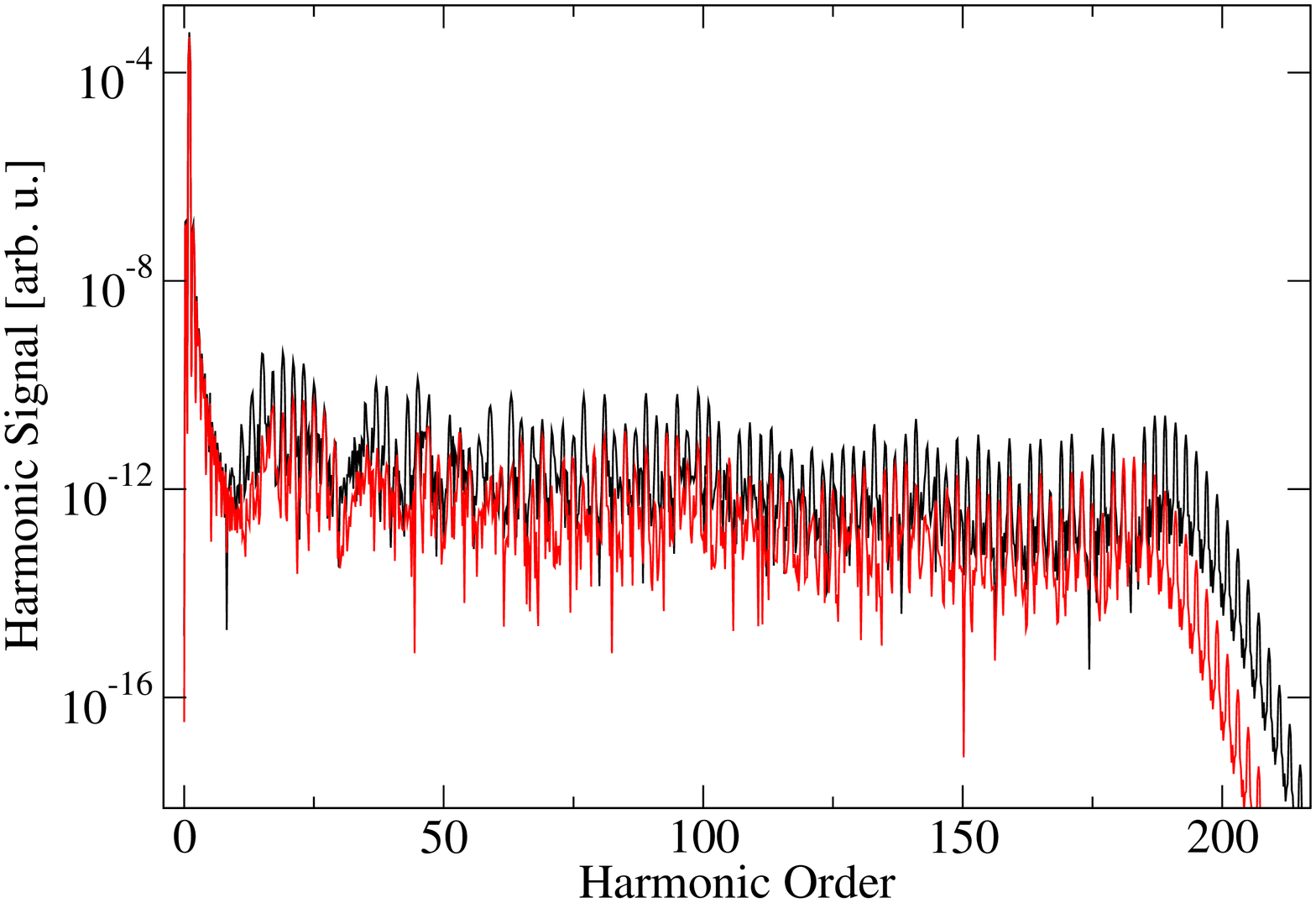}
\caption{(color online). HHG spectra calculated with the soft-core potential (\ref{SC}). The black and grey (red) lines represent the spectra for muonic hydrogen and deuterium, respectively. (a) The laser parameters are $I=1.05\times10^{23}$ W/cm$^{2}$ and $\hslash \omega=59$ eV, corresponding to $\xi\approx 0.04$. 
(b) Same as (a) but for the reduced laser intensity $I=1.05\times10^{21}$ W/cm$^{2}$.
(c) Same as (a) but for the doubled laser frequency $\hslash \omega=118$\,eV.
Note that (a) has been reproduced from Ref.\,\cite{ShMuStBuKe2007} in order to facilitate the comparison between the various spectra.}
\label{fig1a}
\end{figure}

First of all let us consider muonic hydrogen isotopes exposed to very intense VUV laser fields.
In Fig.\,\ref{fig1a}(a) we show the harmonic spectra for muonic hydrogen (where the nucleus is a proton) and muonic deuterium. For muonic hydrogen the spectrum extends further including 60 more harmonics as compared with that of deuterium. The difference of cutoff positions can be understood by inspection of the formula for the spectral cutoff energy, $\epsilon_{\rm max}=I_{p}+3.17U_{p}$ \cite{EbBook1992}. In the present case the ponderomotive energy is given by [see Eq.\,(\ref{RSE})]
\begin{equation}
\label{Up}
U_{p}=\dfrac{e^{2}E^{2}}{4\omega^{2}m_{r}}=\dfrac{e^{2}E^{2}}{4\omega^{2}}\left(\dfrac{1}{m_{\mu}}+\dfrac{1}{m_{n}}\right)
\end{equation}
and is, thus, the larger the smaller the reduced mass is. Consequently, in an intense laser field with $U_p\gg I_p$, muonic hydrogen (H) gives rise to a larger cutoff energy than muonic deuterium (D). The relative difference is about 5\% according to $\epsilon_{\rm max}^{\rm (H)}/\epsilon_{\rm max}^{\rm (D)}\approx m_r^{\rm (D)}/m_r^{\rm (H)}\approx 1.05$. Note that $m_r^{\rm (H)}\approx 0.90m_\mu$, whereas $m_r^{\rm (D)}\approx 0.95m_\mu$.

The nuclear mass effect can also be understood more intuitively within the two-particle picture, instead of the relative motion. The right-hand side of Eq.\,(\ref{Up}) describes a ponderomotive energy that consists of two parts: one for the recolliding muon and one for the recolliding nucleus. The total pondermotive energy is thus a sum of the ponderomotive energies of the muon and the nucleus. Both particles are driven into opposite directions by the laser field and when they recollide, their kinetic energies add up. In this picture the higher cutoff energy of the hydrogen atom arises from the larger ponderomotive energy of the proton as compared to the heavier deuteron \cite{HeHaKe2004}. 

In the opposite situation when $U_{p}\ll I_{p}$, the order of the spectral cutoff positions is reversed, as shown in Fig.\,\ref{fig1a}(b). Here we assumed a laser field with the same frequency as in Fig.\,\ref{fig1a}(a) but a
100 times reduced intensity. The spectra in Fig.\,\ref{fig1a}(b) exhibit a perturbative decay of the signal strength with increasing harmonic order, followed by a large peak around the 35th harmonic. The peak is due to a multiphoton resonance with the first excited atomic state. For muonic hydrogen, the transition energy to this level is 2.00\,keV in the softcore binding potential (\ref{SC}) \cite{EbBook1992}, whereas it amounts to 2.10\,keV for muonic deuterium. When scaled to the laser photon energy, the peaks arise at the corresponding, slightly different positions.

For the parameters assumed in Fig.\,\ref{fig1a}(a), the ratio $\hslash\omega/I_p\approx 0.024$ is relatively low. In fact, the same ratio is obtained when an ordinary hydrogen atom interacts with a mid-infrared laser field of frequency $\hslash\omega = 0.33$\,eV (i.e. wavelength $\lambda=3.7\,\mu$m). Under such circumstances the harmonic signal strength is significantly reduced as compared to HHG in optical or near-infrared fields \cite{MaFr2007,Sc2007,Burgdorfer}. For this reason we provide in Fig.\,\ref{fig1a}(c) the results for a higher driving frequency. As compared to the value employed in Fig.\,\ref{fig1a}(a), the laser frequency has been doubled leading to an increase of the harmonic signal strength by 1-2 orders of magnitude. Since the harmonic order at the cutoff position is proportional to $\omega^{-3}$, the range of emitted harmonics is reduced. At the doubled laser frequency in Fig.\,\ref{fig1a}(c), the different cutoff positions for muonic hydrogen vs. deuterium due to the nuclear mass effect are still clearly visible. However, at the tripled laser frequency (not shown) the cutoff positions practically coincide because the cutoff energies from muonic hydrogen and deuterium differ by less than a photon energy. We point out that the spectra can be distinguished nevertheless by the plateau height which remains substantially higher for muonic hydrogen [see Figs.\,\ref{fig1a}(a) and \ref{fig1a}(c)].

\begin{figure}
\includegraphics*[width=0.64\linewidth,angle=270]{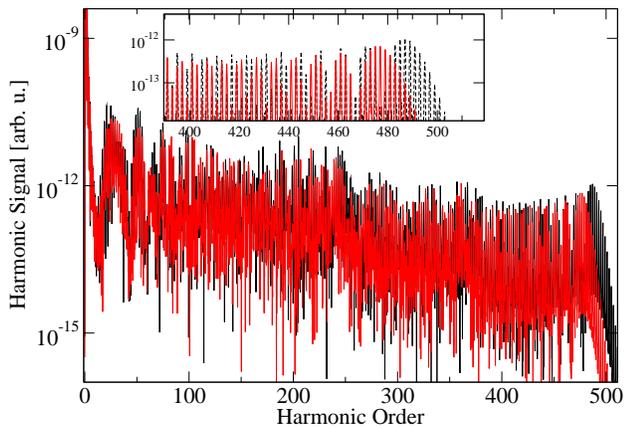}
\caption{(color online). HHG spectra calculated with the soft-core potential (\ref{SC}). The black and grey (red) lines represent the spectrum for muonic $^{3}$He and $^{4}$He, respectively. The laser parameters are $I=8\times10^{24}$ W/cm$^{2}$ and $\hslash \omega=347$ eV. The inset shows an enlargement of the cutoff region.}
\label{fig1d}
\end{figure}

In Fig.\,\ref{fig1d} the harmonic spectra for different isotopes of muonic helium in an ultra-intense XUV laser field are shown. We see that for $^{3}$He the spectrum extends slightly further including 10 more harmonics in contrast with that of $^{4}$He. The difference in the cutoff positions cannot fully be explained by the reduced muon mass here. Formula (\ref{Up}), which holds for hydrogen atoms ($Z=1$), would predict a difference of only 5 harmonics for this case. In the general case ($Z\ge 1$), however, the ponderomotive energy of the relative motion reads
\begin{equation}
\label{Upr}
U^{(r)}_{p}=\dfrac{q_{e}^{2}E^{2}}{4\omega^{2}m_{r}}=\dfrac{e^{2}E^{2}m_{r}}{4\omega^{2}} \left(\dfrac{Z}{m_{n}}+\dfrac{1}{m_{\mu}}\right)^{2}.
\end{equation}
This formula correctly predicts the difference in the cutoff positions in Fig.\,\ref{fig1d}. The effective charge can thus have a measurable impact on the HHG response. 

It is interesting to observe that Eq.\,(\ref{Upr}), in contrast to Eq.\,(\ref{Up}), does not simply separate into a sum of the ponderomotive energies of the muon and the nucleus; i.e., it is different from $U^{\prime}_{p}\equiv (e^2E^2/4\omega^2)(Z^2/m_n+1/m_\mu)$. The reason is that in the case $Z>1$, the center-of-mass does not stay at rest. Rather, the ponderomotive energy $U^{\rm (cm)}_{p}\equiv (Z-1)^2e^2E^2/4\omega^2M$ is connected with its motion. The relation between the various ponderomotive energies is 
\begin{equation}
U^{\prime}_{p}=U^{(r)}_{p} + U^{\rm (cm)}_{p}.
\end{equation}
Only in the case of hydrogen isotopes ($Z=1$) the center-of-mass coordinate remains at rest since the total charge is zero, so that $U^{\rm (cm)}_{p}=0$ and $U^{(r)}_{p}$ fully accomodates the single-particle ponderomotive energies.

%
\subsection{Nuclear size effects}

In this section we investigate the influence of the nuclear extension on the radiation spectra by making use of the hardcore potential (\ref{HC}). For the radial parameter $R$ in Eq.\,(\ref{HC}), the respective rms radius from Table\,\ref{table} is employed. We again consider low-$Z$ isotopes where the largest relative size differences are found. Two isotopes with different radius also differ in mass. We wish to separate, though, the impact of the nuclear size from the nuclear mass effect which was discussed in the previous section. To this end, we compare in the following HHG spectra from different isotopes where the nuclear mass effect has been removed by a suitable adjustment of the laser frequencies and intensities. In accordance with the scaling relations (\ref{scaling}), this was achieved by applying the scaled parameters
$\omega \propto m_{r}$ and $E\propto m_{r}^{2}/q_e$ (at a given value of $Z$). In this manner, the laser-driven muonic isotopes become equivalent to the same ordinary hydrogen atom, with the only difference being the size of the binding nucleus. In particular, the harmonic cutoff positions are forced to coincide this way.

\begin{figure}
(a)\includegraphics*[width=0.64\linewidth,angle=270]{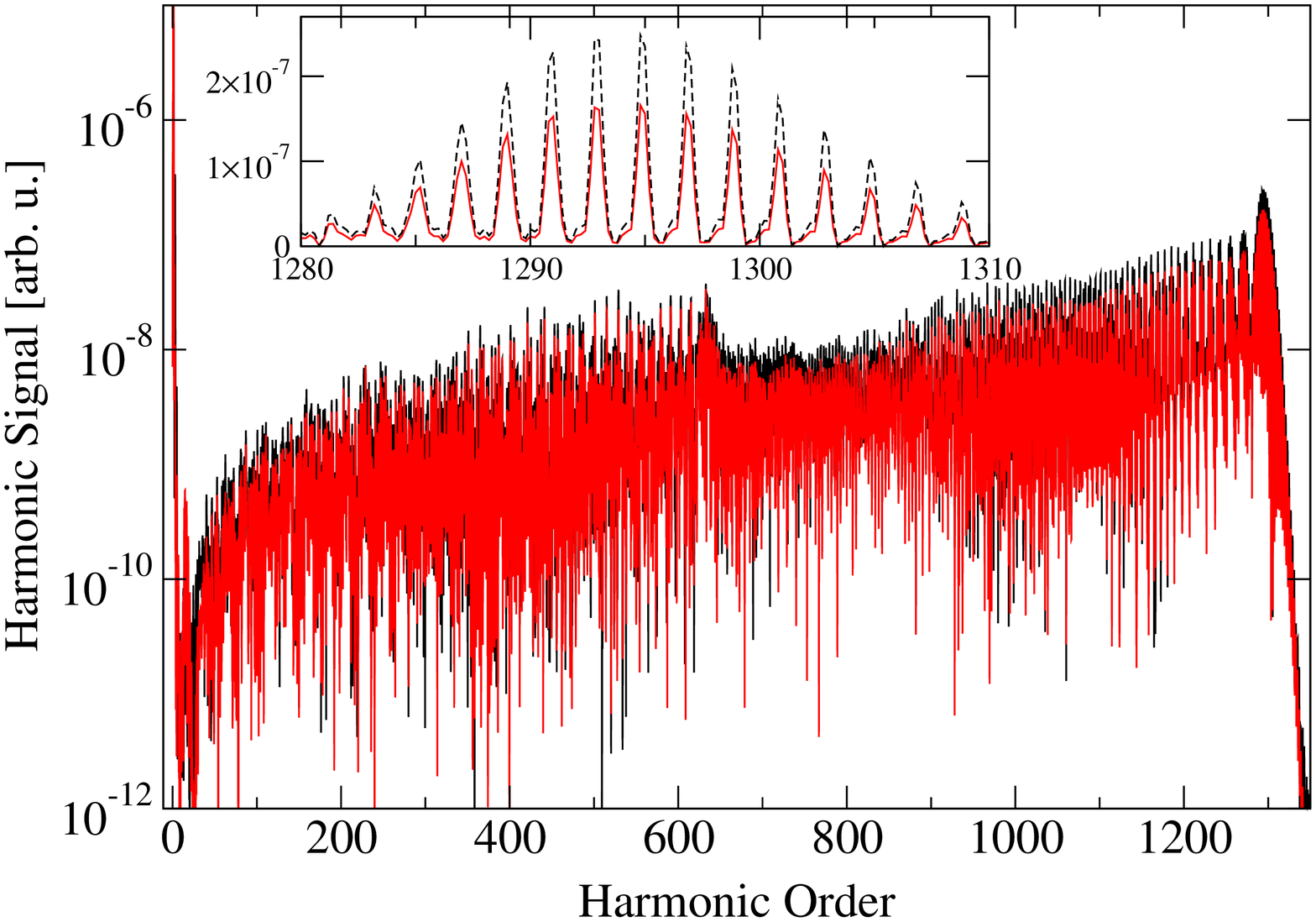}
(b)\includegraphics*[width=0.64\linewidth,angle=270]{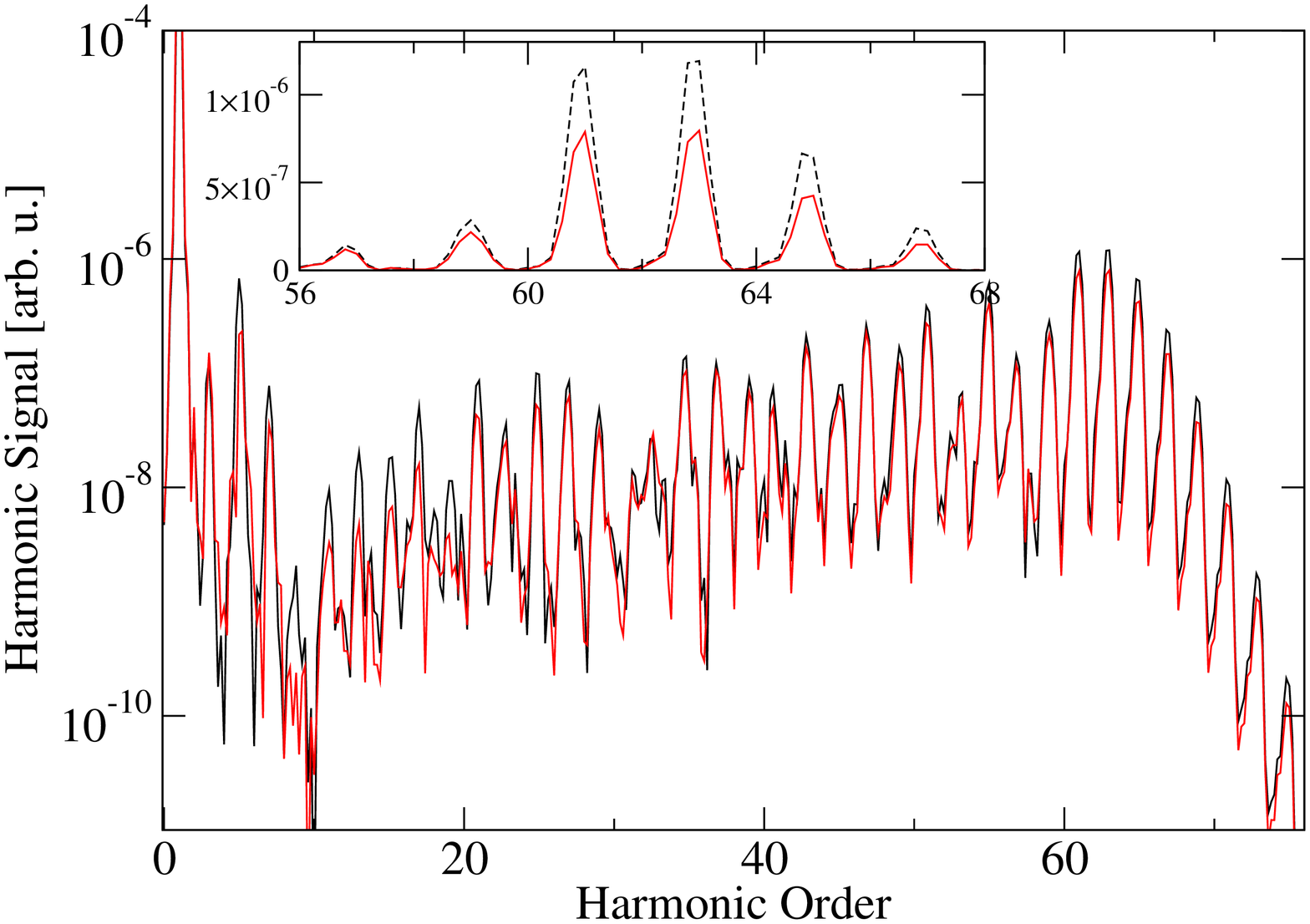} (c)\includegraphics*[width=0.64\linewidth,angle=270]{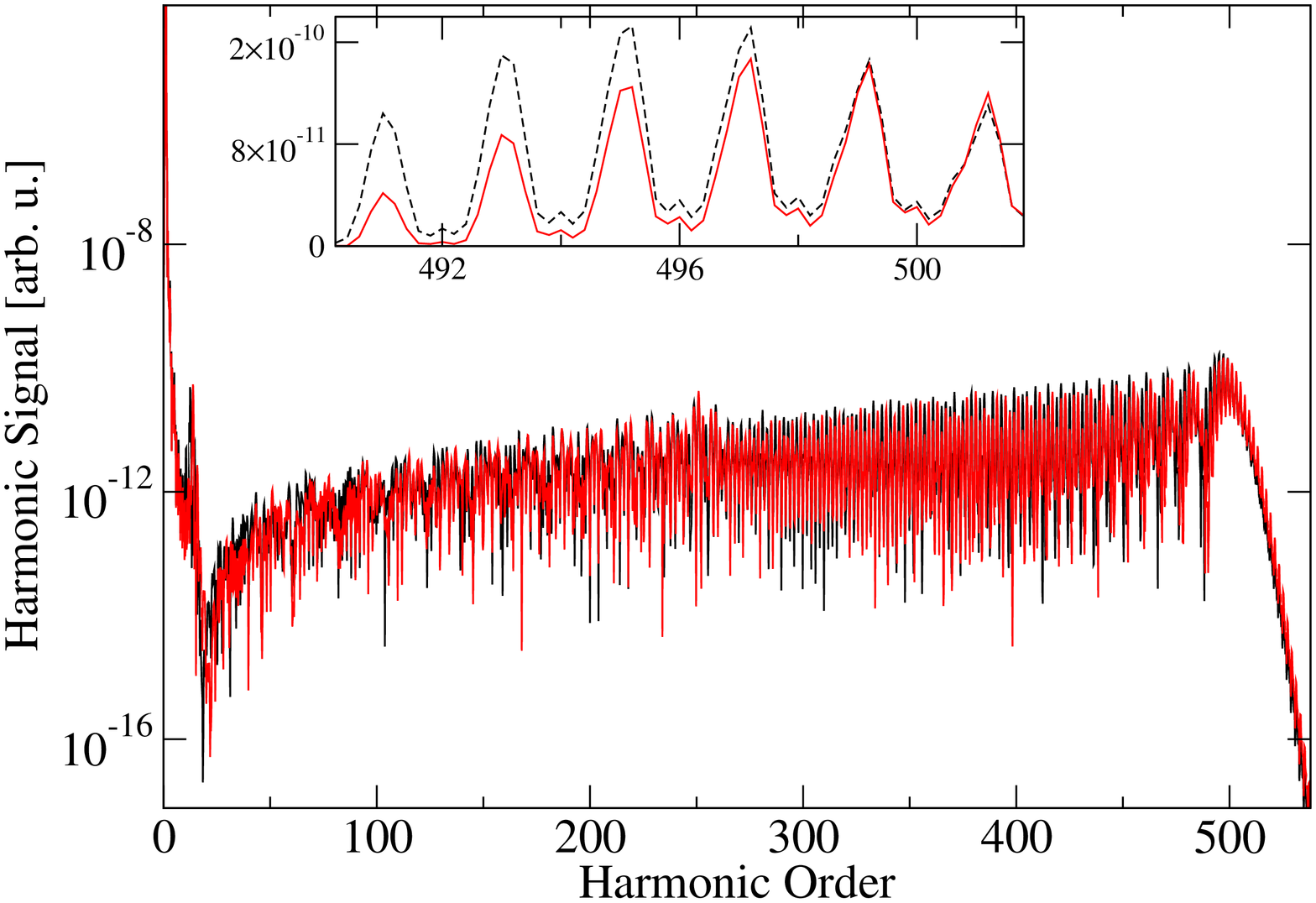}
\caption{(color online). HHG spectra calculated with the hard-core potential (\ref{HC}). (a) The black line represents the spectrum for muonic hydrogen at the laser parameters $I^{\rm (H)}=1.05\times10^{23}$\,W/cm$^{2}$ and $\hslash \omega^{\rm (H)}=59$\,eV. The grey (red) line represents the spectrum for muonic deuterium at the appropriatly scaled laser parameters $I^{\rm (D)}=1.30\times10^{23}$\,W/cm$^{2}$ and $\hslash \omega^{\rm (D)}=62$\,eV in order to compensate for the nuclear mass effect. The inset shows an enlargement of the cutoff region on a linear scale (reproduced from \cite{ShMuStBuKe2007} to allow for a direct comparison).
(b) Same as (a) but for the laser parameters $I^{\rm (H)}=1.05\times10^{23}$\,W/cm$^{2}$, $\hslash \omega^{\rm (H)}=177$\,eV and $I^{\rm (D)}=1.30\times10^{23}$\,W/cm$^{2}$, $\hslash \omega^{\rm (D)}=186$\,eV, whereas (c) refers to $I^{\rm (H)}=3.8\times10^{22}$\,W/cm$^{2}$, $\hslash \omega^{\rm (H)}=59$\,eV and $I^{\rm (D)}=4.68\times10^{22}$\,W/cm$^{2}$, $\hslash \omega^{\rm (D)}=62$\,eV.}
\label{fig3a}
\end{figure}

We start with the calculations for muonic hydrogen and deuterium shown by Fig.\,\ref{fig3a}. As mentioned above, in order to avoid residual signatures from the nuclear mass effect we apply in Fig.\,\ref{fig3a}(a) the laser parameters $I^{\rm (H)}=1.05\times10^{23}$\,W/cm$^{2}$ and $\hslash \omega^{\rm (H)}=59$\,eV to muonic hydrogen, whereas muonic deuterium is subject to the parameters $I^{\rm (D)}=1.30\times10^{23}$\,W/cm$^{2}$ and $\hslash \omega^{\rm (D)}=62$\,eV. Regarding the overall shape of both spectra we observe a dip at low harmonics around the order $n \sim I_{p}/\hslash\omega$, followed by a rising plateau region. These common features are due to the hardcore potential used, in contrast to the softcore potential (\ref{SC}). In fact, a similar behaviour of the harmonic response was found in \cite{HuStBeSaMi2002} and attributed to so-called non-tunneling harmonics in very steep potentials. Our main focus, however, lies on the relative difference between the two spectra. The harmonic signal from muonic hydrogen is larger (by about 50\,\% in the cutoff region) than that from muonic deuterium. The reason is that in the case of muonic hydrogen the nuclear radius is smaller which generates a steeper potential near the origin [see Eq.\,(\ref{HC})]. This leads to a larger potential gradient in this region which accelerates the atomic dipole according to Ehrenfest's theorem \cite{GoSaKa2005}. The muon in hydrogen is thus more strongly accelerated, leading to enhanced harmonic emission. 

When the laser frequency is enhanced three times as compared to Fig.\,\ref{fig3a}(a) while the laser intensity is kept constant, the relative difference between the HHG spectra from muonic hydrogen and deuterium stays approximately constant. As shown in Fig.\,\ref{fig3a}(b), the signal from muonic hydrogen remains larger by about 50\,\% in the cutoff region.
When instead the frequency of Fig.\,\ref{fig3a}(a) is kept but the field intensity is decreased down to $3.8\times10^{22}$ W/cm$^{2}$ ($4.68\times10^{22}$ W/cm$^{2}$) for muonic hydrogen (deuterium), then the harmonic signals only differ by about 10\,\% in the cutoff region, as shown in Fig.\,\ref{fig3a}(c).

\begin{figure}
(a)\includegraphics*[width=0.64\linewidth,angle=270]{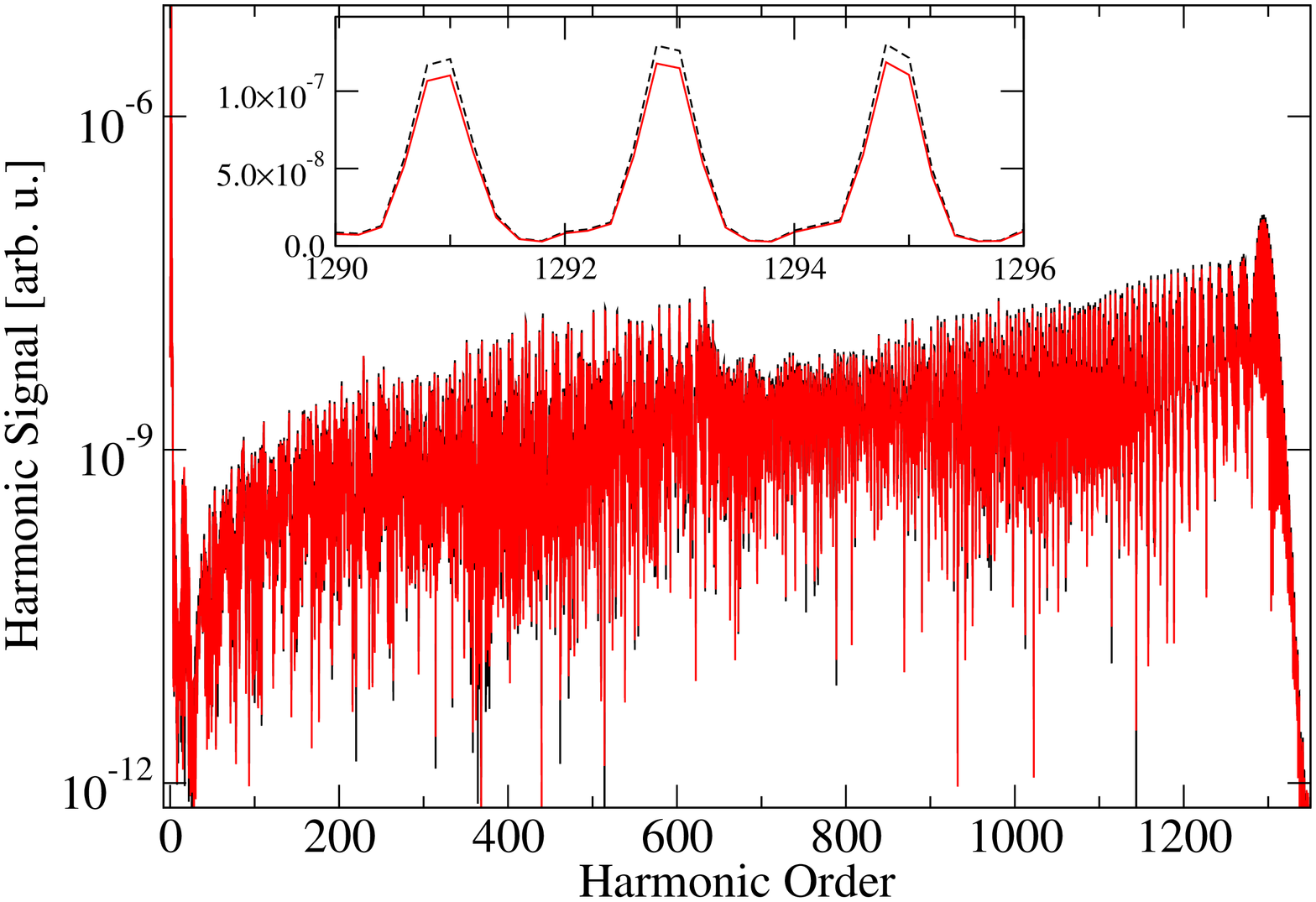} (b)\includegraphics*[width=0.64\linewidth,angle=270]{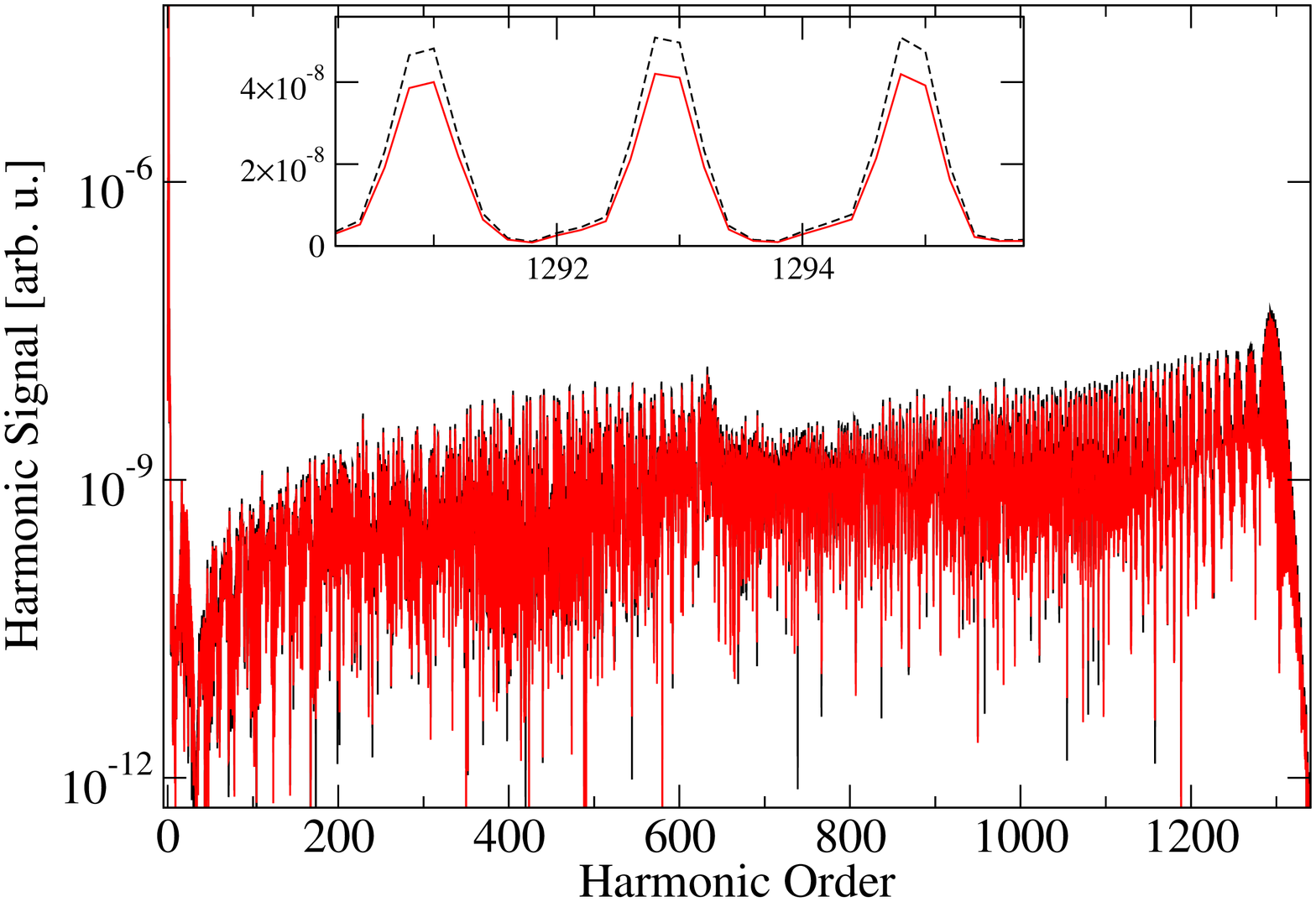}
\caption{(color online). HHG spectra calculated with the hard-core potential (\ref{HC}). (a) The black line shows the spectrum for muonic $^{4}$He at the laser parameters $I^{\rm (^4He)}=8.7\times10^{24}$\,W/cm$^{2}$ and $\hslash \omega^{\rm (^4He)}=255$\,eV. The grey (red) line represents the spectrum for muonic $^{3}$He at the accordingly scaled values $I^{\rm (^3He)}=8.3\times10^{24}$\,W/cm$^{2}$ and $\hslash \omega^{\rm (^3He)}=253$\,eV. (b) Same as (a) but for muonic $^{9}$Li at $I^{\rm (^9Li)}=1.06\times10^{26}$\,W/cm$^{2}$, $\hslash \omega^{\rm (^9Li)}=583$\,eV [black line] and muonic $^{6}$Li at $I^{\rm (^6Li)}=1.01\times10^{26}$\,W/cm$^{2}$, $\hslash \omega^{\rm (^6Li)}=580$\,eV [grey (red) line].}
\label{fig3d}
\end{figure}

In the case of muonic helium isotopes, the harmonic signal from $^{4}$He is expected to be larger than that from $^{3}$He. For $^{4}$He, though containing one more neutron, is a doubly magic nucleus of very compact size (see Table\,\ref{table}). This expectation is confirmed by Fig.\,\ref{fig3d}(a), where a relative difference of about 10\,\% in the cutoff region is observed. The fact that the difference is reduced in comparison with the muonic hydrogen isotopes can be attributed to the smaller relative difference of the nuclear radii: $R^{\rm (^3He)}/R^{\rm (^4He)}\approx 1.16$, whereas $R^{\rm (D)}/R^{\rm (H)}\approx 2.44$. On the other hand, however, the muon comes closer to the binding nucleus when the charge number $Z$ increases which should enhance the sensitivity to the nuclear size. This circumstance becomes important when we move on to muonic lithium ($Z=3$). Here, the difference between the harmonic signals from muonic $^{6}$Li versus muonic $^{9}$Li amounts to about 20\,\% in the cutoff region [see Fig.\,\ref{fig3d}(b)] which is larger than for the helium isotopes although the ratio of the nuclear radii $R^{\rm (^6Li)}/R^{\rm (^9Li)}\approx 1.14$ is similar here. In order to facilitate the comparison between helium and lithium, the laser parameters in Fig.\,\ref{fig3d} were chosen to generate a uniform cutoff position.
Finally, a comparison of the radiative responses from muonic $^{20}$Ne and $^{23}$Ne at the corresponding field parameters reveals almost identical HHG spectra (not shown). The relative enhancement of the HHG signal from the smaller isotope $^{23}$Ne as compared with $^{20}$Ne is of the order of 1\%. Here, the ratio of the nuclear radii, $R^{\rm (^{20}Ne)}/R^{\rm (^{23}Ne)}\approx 1.03$, is close to unity. We note moreover that in the cases of lithium and neon practically no mass effect needs to be compensated since the corresponding reduced muon masses almost coincide.

\begin{figure}
\includegraphics*[width=0.94\linewidth,angle=0]{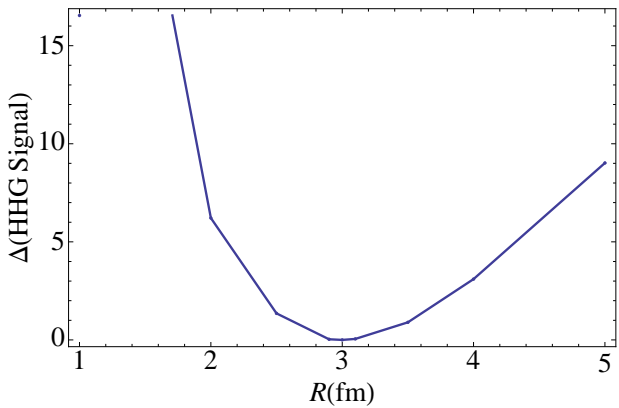}
\caption{Dependence of the harmonic signal strength at the cutoff on the radius of the neon-like nucleus to which the muon is bound. The signal strength is expressed as the normalized difference $\Delta\equiv [S(R)-S(R_0)]/S(R_0)$, where $S(R)$ denotes the near-cutoff signal strength for nuclear radius $R$ and $R_0=3$\,fm is the reference value. We point out that the nuclear size variation shown here largely extends beyond the Ne isotopes which exist in nature where always $R\approx3$\,fm \cite{An2004}.}
\label{fig4}
\end{figure}

So far we have restricted our investigation of the nuclear size effects to isotopes which exist in nature. It is instructive to extend the consideration by varying the nuclear radius artificially in order to gain a more complete picture of the physical mechanisms involved. Considering hydrogenic nuclei with charge number $Z=1$ and radius $R$ increasing from 1\,fm to 5\,fm, we obtain a monotonously decreasing HHG signal, in accordance with the explanation given above regarding Fig.\,\ref{fig3a}. For neon-like nuclei with $Z=10$, however, a minimum harmonic emission strength arises around $R\approx 3$\,fm. When the nuclear extension is artificially enhanced further to $R=5$\,fm, the HHG signal is growing again (see Fig.\,\ref{fig4}). This different behavior can be attributed to the atomic ionization potential $I_p$ which significantly decreases in the neon-like case when $R$ in Eq.\,(\ref{HC}) is increased. Note that the atomic Bohr radius of hydrogenlike muonic neon amounts to 25\,fm only, rendering this system more sensitive to strong deviations from a point nucleus. Hence, the probability for tunneling ionization in the laser field, being the first step of HHG, is enhanced which leads to amplified harmonic emission for $R>3$\,fm.

When considering laser-driven recollisions, the de-Broglie wavelength of the returning quantum wave packet can be of importance as well. In fact, for the case of ordinary molecules it has been shown that the electron wavelength can become as small as the internuclear distance within the molecule, causing characteristic diffraction patterns \cite{diffrac}. In the present case of muonic atoms, a similar effect could in principle arise when the de-Broglie wavelength of the recolliding muon compares with the nuclear size. However, the wavelength of a muon with a kinetic energy of a few MeV (see Table\,\ref{tableCutoff}) amounts to about 50\,fm which exceeds any nuclear radius substantially and thus prevents diffractive muon-nucleus scattering.

Concluding this section, we have shown that the plateau height of HHG spectra from muonic atoms is sensitive to the finite nuclear size. Smaller nuclei within the range of existing isotopes lead to enhanced harmonic emission. We note, however, that -- contrary to the nuclear mass effect in Sec.\,III.B -- the influence of the nuclear size might be overestimated by our 1D hardcore-potential approach as the muon meets the nucleus more often than in the real 3D case. A calculation in higher dimensionality could provide more accurate quantitative predictions on the nuclear size effect whose physical origin and basic features have been presented here.

%
\subsection{Comparison with electronic systems}
Finite nuclear size effects -- in the absence of any external laser field -- have been revealed in high-precision spectroscopy of electron transitions in (ordinary) highly charged ions (see, e.g., \cite{Bruhns,Brandau2008} for recent experiments). When such ionic systems are exposed to a superintense laser field \cite{HHG-HCI}, nuclear signatures may be present in their high-harmonic response as well. It is of interest to compare the expected effects with those found for laser-driven muonic atoms in Sec. III.C. To this end we perform a simple analysis which is based on nonrelativistic Schr\"{o}dinger theory; the relativistic electron motion in highly charged ions is ignored in this rather qualitative discussion. 

We assume a hydrogenlike system of nuclear charge number $Z$ and employ the mass scaling parameter $\rho$ from Eq. (5), with $\rho\approx 1/200$ for a muonic atom and $\rho=1$ for an electronic ion. The K-shell Bohr radius, binding energy, and Coulombic field strength amount to $a_K(Z,\rho)=a_0\rho/Z$, $I_p(Z,\rho)=\epsilon_0 Z^{2}/\rho$, and $E_K(Z, \rho)=E_0 Z^{3}/\rho^{2}$, respectively, where $a_0$, $\epsilon_0$ and $E_0$ denote the corresponding quantities for ordinary hydrogen. The nuclear radius can be approximated roughly as $R(Z)\approx 1.2 (2Z)^{1/3}\,{\rm fm}$ and has a typical relative variation among different isotopes of a few percent (except for hydrogen vs. deuterium). Similar finite nuclear size effects in the HHG spectra can be expected when the ratio $R(Z)/a_K(Z,\rho)\propto Z^{4/3}/\rho$ has a similar value for two atomic systems that are compared. This is the case, e.g., for electronic U$^{91+}$ (where $Z=92, \rho=1$) and muonic He$^{+}$ (where $Z=2, \rho\approx1/200$). 

The above relations imply, however, that the binding energy and electric field strength in the electronic ion are substantially larger than in the muonic atom when both have the same ratio of $Z^{4/3}/\rho = const$. As a consequence, the laser frequency and intensity that must be applied to the electronic highly charged ion in order to reveal finite nuclear size effects in the harmonic response, need to be larger than in the muonic atom case. Against this background, muonic atoms appear as more favorable systems than ordinary heavy ions to study the influence of the nuclear size on the HHG process.

%
\section{Conclusion and Outlook}
Motivated by the sustained progress in the development of powerful laser sources, we have studied the harmonic radiation which is emitted by muonic atoms exposed to high-intensity, high-frequency laser fields. It was shown that maximum harmonic cutoff energies in the MeV domain can be achieved, rendering this species of exotic atoms promising candidates for the generation of (weak) ultrashort coherent $\gamma$-ray pulses which might be employed to trigger photo-nuclear reactions. Our results demonstrate moreover that strongly laser-driven muonic atoms can, in principle, be utilized to dynamically gain structure information on nuclear ground states via their high-harmonic response. (1) On the one hand, the harmonic cutoff position extends to larger values for isotopes of smaller mass. For the hydrogen isotopes this effect is fully explicable in terms of the reduced mass, whereas for atomic numbers $Z>1$ an additional contribution stems from an effective muon charge which affects the relative motion generating the harmonics. (2) On the other hand, the harmonic signal strength additionally depends on the nuclear size, being enhanced for more compact isotopes. Corresponding nuclear size effects in the high-harmonic emission from ordinary highly-charged ions are expected to be less pronounced.

Furthermore, we point out that the interaction of a muonic atom with ultrastrong laser fields may lead to excitation of the nucleus. Nonresonant nuclear Coulomb excitation has recently been studied when a bound muonic wave packet is driven into coherent oscillations by an external laser field \cite{NPA}; the resulting nuclear excitation probabilities were found to be small, though. When the laser field is sufficiently strong to ionize the muon as in the HHG scenario, however, the kinetic energy gain in the continuum up to the MeV range allows for nuclear excitation upon the muon-nucleus recollision. Corresponding studies of laser-driven electron-impact excitation of the nucleus have been carried out in ordinary atoms and ions \cite{recollision}. It is even conceivable to conduct pump-probe experiments on excited nuclear levels: the periodically driven muon can first excite the nucleus and then probe the excited state and its deexcitation mechanism during a subsequent encounter.

%
\begin{acknowledgments}
We are grateful to A. Staudt for valuable discussions and for providing us with his computer code. We also thank A. D. Bandrauk, K. Z. Hatsagortsyan, C. H. Keitel, N. L. Manakov, and J. Ullrich for useful conversations. A.~S. acknowledges support by the Higher Education Commission (HEC), Pakistan and Deutscher Akademischer Austauschdienst (DAAD).
\end{acknowledgments}

%


\begin{thebibliography}{10}
\bibitem{GoSaKa2005} The importance of the analytical behavior of the potential near the origin for the HHG process in ordinary atoms has been studied in Ref.\,\cite{Gordon}.
\end{thebibliography}
\end{document}